\documentclass[twocolumn]{aastex62}
\pdfoutput=1 
\usepackage{amsmath,amstext}
\usepackage[T1]{fontenc}
\usepackage{apjfonts} 
\usepackage[figure,figure*]{hypcap}
\usepackage{tabularx} 
\usepackage{supertabular}

\shortauthors{Greklek-McKeon \& Deming}

\begin{document}

\title{Killing Planet Candidates with EVEREST}
\shorttitle{Killing Planets}

\author{Michael Greklek-McKeon}
\author{Drake Deming}
\affiliation{Department of Astronomy, 4296 Stadium Drive, University of Maryland, College Park, Maryland, 20742, USA}
\correspondingauthor{Michael Greklek-McKeon}
\email{mgreklek@terpmail.umd.edu, ddeming@astro.umd.edu}

\begin{abstract}
We exploit high quality photometry from the EVEREST pipeline to evaluate false-positive exoplanet candidates from the K2 mission. We compare the practical capabilities of EVEREST's pixel-level decorrelation scheme to the data analysis pipelines widely used at the time of these planet candidates' discovery.  Removing stellar variability from the EVEREST-corrected light curves, we search for potential secondary eclipses.  For each object exhibiting a secondary eclipse, we compare the implied brightness temperature of the planet candidate to its calculated equilibrium temperature.  We thereby identify objects whose brightness temperature is too high to be consistent with a planet.  We identify seven systems previously flagged as planetary candidates in preliminary vetting pipelines, and use EVEREST to instead identify six of them as eclipsing binaries.  We also project the importance of optimal photometric vetting for TESS data. We find that the majority of blended eclipsing binaries could be identified using TESS photometry, and a systematic study of that kind could in principle also yield valuable information on the mass ratio distribution in stellar eclipsing binaries.  
\end{abstract}


\section{Introduction}

Kepler's extended mission, K2, experienced systematic pointing errors due to the pseudo-stable fine spacecraft control using the two functioning reaction wheels and the balancing pressure of sunlight on the solar panels.  This mechanism, described in detail in \cite{K2}, yields roll motion that is two orders of magnitude greater than the typical 6 hour pointing errors in the Kepler primary mission as described by \cite{howweroll}. Fortunately, improved data processing pipelines have increased K2's photometric precision and enabled the further discovery of many planet candidates in new fields of observation near the ecliptic plane. Numerous pipelines have been developed that correct the systematics present in the raw K2 photometry, including: \textsc{k2sff} \citep{VJ14}, \textsc{k2p2} \citep{K2P2}, \textsc{k2sc} \citep{K2SC}, \textsc{k2varcat} \citep{k2varcat,k2varcat2}, \textsc{k2phot} \citep{k2phot}, and \textsc{EVEREST} \citep{everest,everest2}. These pipelines have been largely used in the discovery of the 473 planet candidates attributed to the K2 mission. 234 of these candidates come from \cite{vanderburg}, hereafter referred to as V16, which identifies the 234 planetary candidates around 208 stars. The pipeline used to correct for systematics in V16 is based upon \textsc{k2sff}. It operates by completely removing data points collected during thruster firing, and correlates the flux of a bright star in the field to the centroid position of the target, and then subsequently removes this trend to account for the roll motion drift of the spacecraft. Of the 234 candidates identified by V16, several overlap with the systems in \cite{crossfield}, hereafter referred to as C16, which identifies 197 candidates from K2's first five fields. 

Vetting exoplanet candidates often requires high resolution imaging and radial velocity measurements; however, the total vetting effort can be minimized if the space-borne photometry can be pushed to the highest possible precision.  High precision photometry can help to identify false positives in planet candidate samples, for example by revealing the presence of significant secondary eclipses. Of the many pipelines that have been used to process K2 data, some are specific to the studies that utilize them, while others have been created for use by the community. Several individualized pipelines, like that of C16, V16, and also that of the general \textsc{k2p2} pipeline described in \cite{K2P2}, are variations built upon the 1D self-flat-fielding (SFF) technique introduced by \cite{VJ14}. The EVEREST pipeline also utilizes an SFF principle, but is more broadly based on pixel-level decorrelation (PLD, \citealp{deming}). EVEREST extends the PLD originally developed for correcting systematics in \textit{Spitzer} observations by \cite{deming} up to third order terms in the pixel fluxes. This main difference, among other smaller factors, sets EVEREST apart from the other K2 photometric pipelines. As demonstrated in \cite{everest}, EVEREST achieves the highest average precision of any of these pipelines for unsaturated K2 stars, and it is being used with increasing frequency by K2 investigators.

In this paper, we explore to what degree the best possible photometry can identify and characterize false positives through photometry alone, and we demonstrate the potential of EVEREST as an important first step to validate exoplanets in the K2 data, and potentially in the TESS data.  We specifically build on the suspicions of the previous works by \cite{vanderburg} and \cite{armstrong} to reveal seven planet candidates as being likely blended eclipsing binaries (BEBs, e.g. \citealp{collins}), by showing that they have clear secondary eclipses, whose depth is inconsistent with a planetary nature.  We then estimate stellar parameters for these eclipsing systems.

\section{Selection of Candidates}
We reviewed systems listed in the papers that produced the two largest sets of planet candidates from the K2 mission, V16 and C16. These papers cover only the first five fields observed by K2. We examine seven systems currently listed as planetary candidates, and classify six of them instead as false positive eclipsing binaries. The systems examined are: EPIC 202071289, 202071635, 202072596, 203867512, 205148699, 206135267, and 203753577.

All seven systems appear in V16 as planetary candidates. 25 of the candidates in that work are noted as "deep transits", which have depths larger than ~5\%. V16 acknowledges that systems with transit depths this large are very likely to be eclipsing binary false positives, but nevertheless includes them as planet candidates because they pass the vetting pipeline. EPIC 206135267 is included in this list of "deep transits". Additionally, four of these systems appear in C16: EPIC 206135267, 202071289, 205148699, and 203867512. C16 notes that EPIC 206135267 is almost certainly an eclipsing binary because the primary transit flux drops to less than 50\% of its total value, but still lists it as a candidate, as it has yet to be formally invalidated (or validated). Likewise, \cite{crossfield} "cannot validate" EPICs 202071289, 205148699, and 203867512, and includes them as planet candidates, but notes that EPIC 205148699 has RV variations in phase with the transit signal and with semi-amplitude $\sim$28km/s, indicating that the system is likely an eclipsing binary, and that EPIC 203867512 has a nearby star, which does not definitively prove that the system is a false positive, but nonetheless deeply complicates any effort to validate the candidate. Finally, \cite{schmitt} observes a companion of EPIC 206135267 with Keck NIRC2 AO imaging.

All of these systems are currently listed in the NASA exoplanet archive as K2 planetary candidates: \href{https://exoplanetarchive.ipac.caltech.edu/cgi-bin/TblView/nph-tblView?app=ExoTbls&config=cumulative&constraint=koi_pdisposition+like+\%27CANDIDATE\%27}{exoplanetarchive.ipac.caltech.edu}. The ability of EVEREST's improved photometry is essential in revealing the non-planetary nature of these objects that were otherwise originally classified as candidates from their respective  pipelines. With EVEREST, these systems and potentially many more can be ruled out as false positives immediately from only photometric data, without having some of them first pass a vetting pipeline only to require further observations attempting to invalidate them with direct imaging or RV observations.

For our examination of these systems previously identified as planet candidates, we use the EVEREST pipeline to obtain reduced photometric data and search for the presence of secondary eclipses. The EVEREST code used to correct the raw data used in our analysis comes from \cite{everest}, which is an open-source pipeline. While we use the period for these systems reported in V16, we use the corrected flux values from the EVEREST pipeline, rather than V16's publicly available corrected light curves of the raw K2 data, which are released at \href{https://www.cfa.harvard.edu/~avanderb/k2.html}{www.cfa.harvard.edu}. While the techniques of \cite{vanderburg} were pioneering, the difference in corrected photometric quality with EVEREST can be significant, as can be seen in the example shown in Figure 1 from EPIC 202072596.

\begin{figure}
  \includegraphics[width=\linewidth]{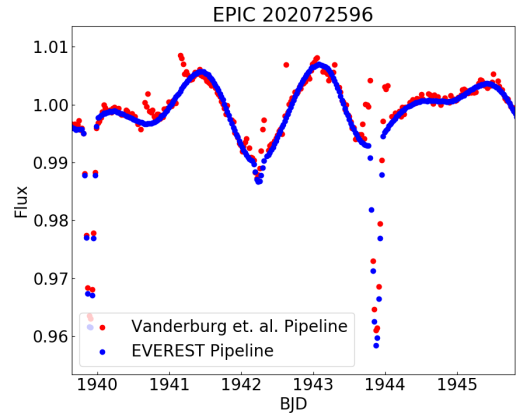}
  \caption{A comparison of the corrected light curves between EVEREST and \cite{vanderburg} for EPIC 202072596, emphasizing the difference in overall scatter between the two.}
  \label{Fig 1}
\end{figure}

\section{Methodology}
We reveal false positives in this dataset by establishing the presence of a secondary eclipse in the phase folded light curve. We examined the systems from every entry in the table of planetary candidates from \cite{vanderburg} that had a planet to star ratio of 5\% or greater, to maximize the chance of discovering a significant secondary eclipse. The flux for each system was initially plotted over the full period of observation using EVEREST and closely visually examined for deep primary transits with potential secondary dips. The non-normalized light curves were then phase folded by the period reported in V16 and examined for any potential secondary eclipses that recur at the same phase. This visual inspection revealed 13 systems with potential secondaries, and after further fitting to remove stellar variability and long term systematic trends unlikely to affect transit signals, the seven systems described in this paper were selected and six of them further determined to be false positives. 

Estimates of the phase and duration of the primary transit and secondary eclipse for each system were made as part of the masking process. The first point of ingress and last point of egress were taken as the first and last point to deviate by 1$\sigma$ or more from the nearby flux, and the mask began and ended at the points immediately preceding and following these two points, respectively. All points within the mask associated with a transit or eclipse were removed from the light curve, so a fit could be optimized to follow only the out of transit flux. The eclipses for the seven systems in this study were symmetric, so the deepest transit and eclipse points coincided with the center of the transit and eclipse masks.

Each system exhibited significant variability in the out-of-transit flux, so after the primary and secondary eclipses were masked, an 8-term Fourier signal was fit to the out of transit flux, and divided out from the complete observed signal to detrend and normalize it. This Fourier fit correction process was iterated 10 consecutive times for each system, to reveal as much of the secondary eclipse as possible from the contaminating background fluctuations in stellar intensity. The resulting corrected, detrended, and phase-folded light curves each showed a clear secondary eclipse of significant depth, indicating the non-planetary nature of the transiting object. We determine the significance of the secondary eclipse by finding the combined significance for the deviations in flux from unity over all the points in the secondary mask, and accept any secondary of 10$\sigma$ significance or greater. The shallowest secondary in the study, from EPIC 203753577, has $\sigma$ > 10.3. The clear secondary eclipses for all systems are shown in Figure 2. To be sure that these secondary eclipse detections were not spurious results of the fits we did to remove the out of transit stellar variation, we performed the same analysis independently on only the first half, and then only the  second half of the dataset. While these data have a higher variance, we can still clearly resolve each of the secondary eclipses, in both cases. The results from only the first half of the data can be seen in Figure 3. 

\begin{figure*}
  \includegraphics[width=\textwidth]{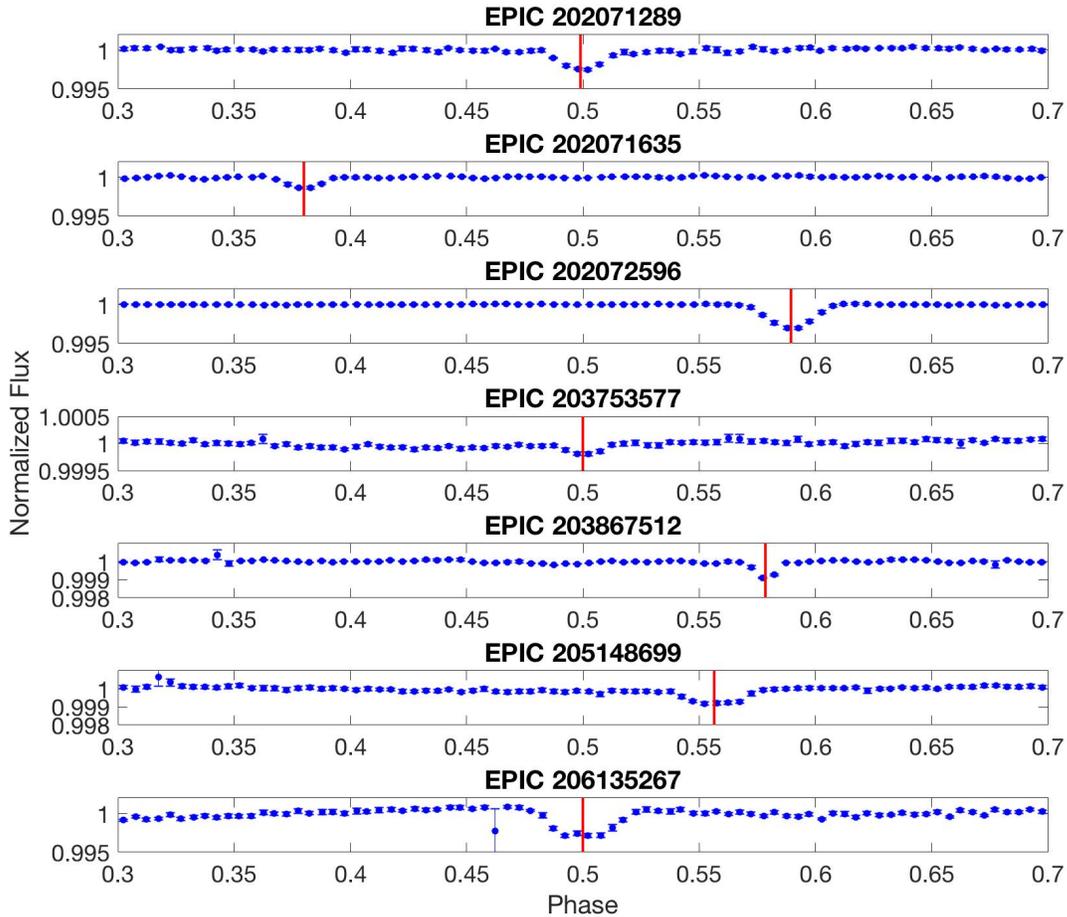}
  \caption{Large secondary eclipses visible in the phase folded light curves for each of the seven systems examined in this study. The phase-folding was done after stellar variability was removed and the flux was normalized and binned at size 0.005. Several systems are eccentric.}
  \label{Fig 2}
\end{figure*}

\begin{figure*}
  \includegraphics[width=\textwidth]{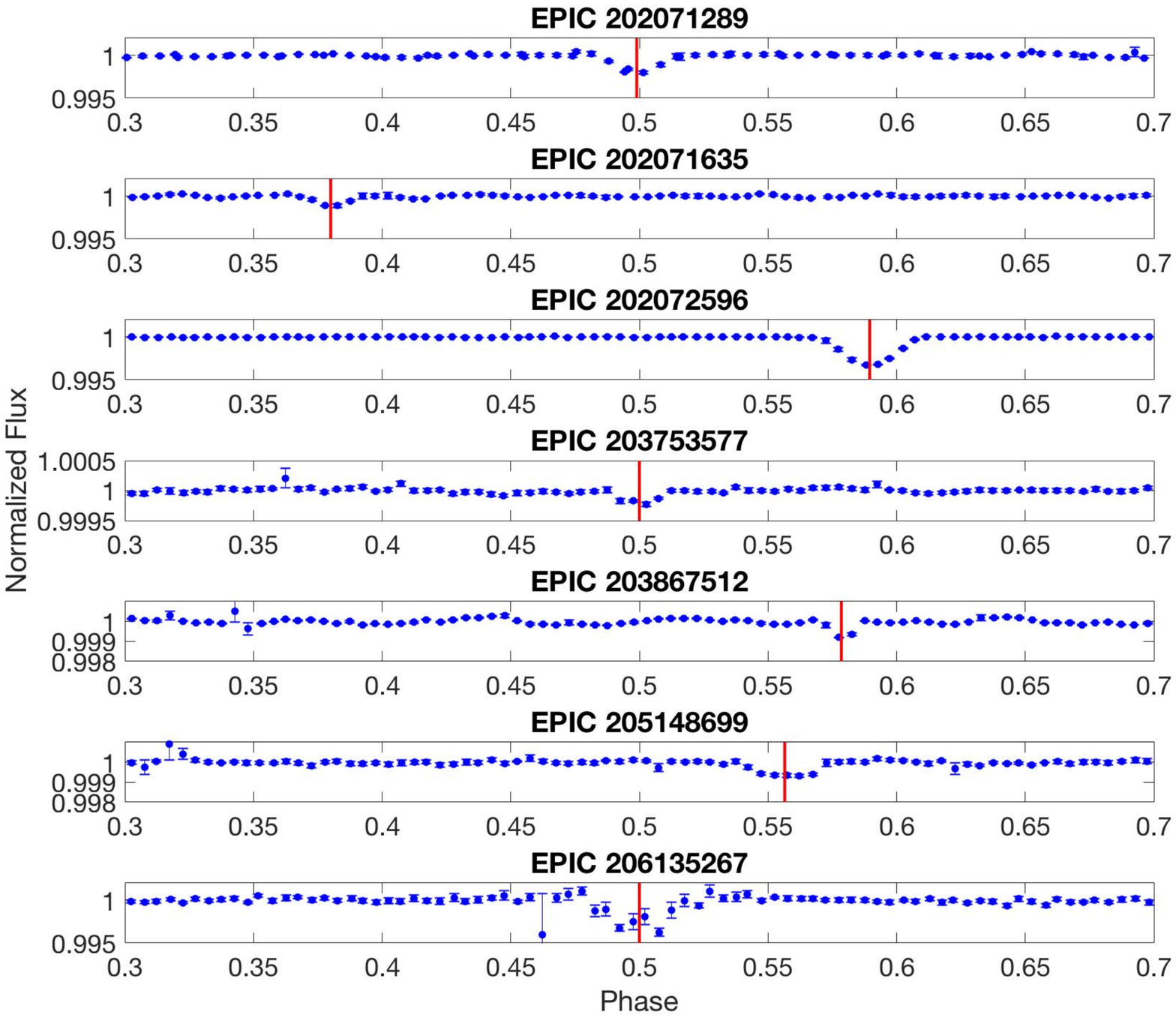}
  \caption{Large secondary eclipses visible in the phase folded light curves, after stellar variability has been removed, using the first half of dataset only, binned with a bin size of 0.005.}
  \label{Fig 3}
\end{figure*}

Some of the systems actually showed a dip in V16's corrected photometry near the phase of the secondary eclipse identified in this work using EVEREST photometry, but the higher variability in the total flux from the V16 pipeline reduction led to greater variance where the secondary occurred. This washed out the significant signal, even after our same out of transit sinusoid fitting process was applied to the corrected light curve. The difference in secondary eclipse quality between EVEREST and V16 for a system showing an especially large improvement can be seen in Figure 4. For the seven systems examined in this work, the average improvement in the mean standard deviation from the V16 photometry to the detrended EVEREST photometry was by a factor of 11.1.

\begin{figure}
  \includegraphics[width=\linewidth]{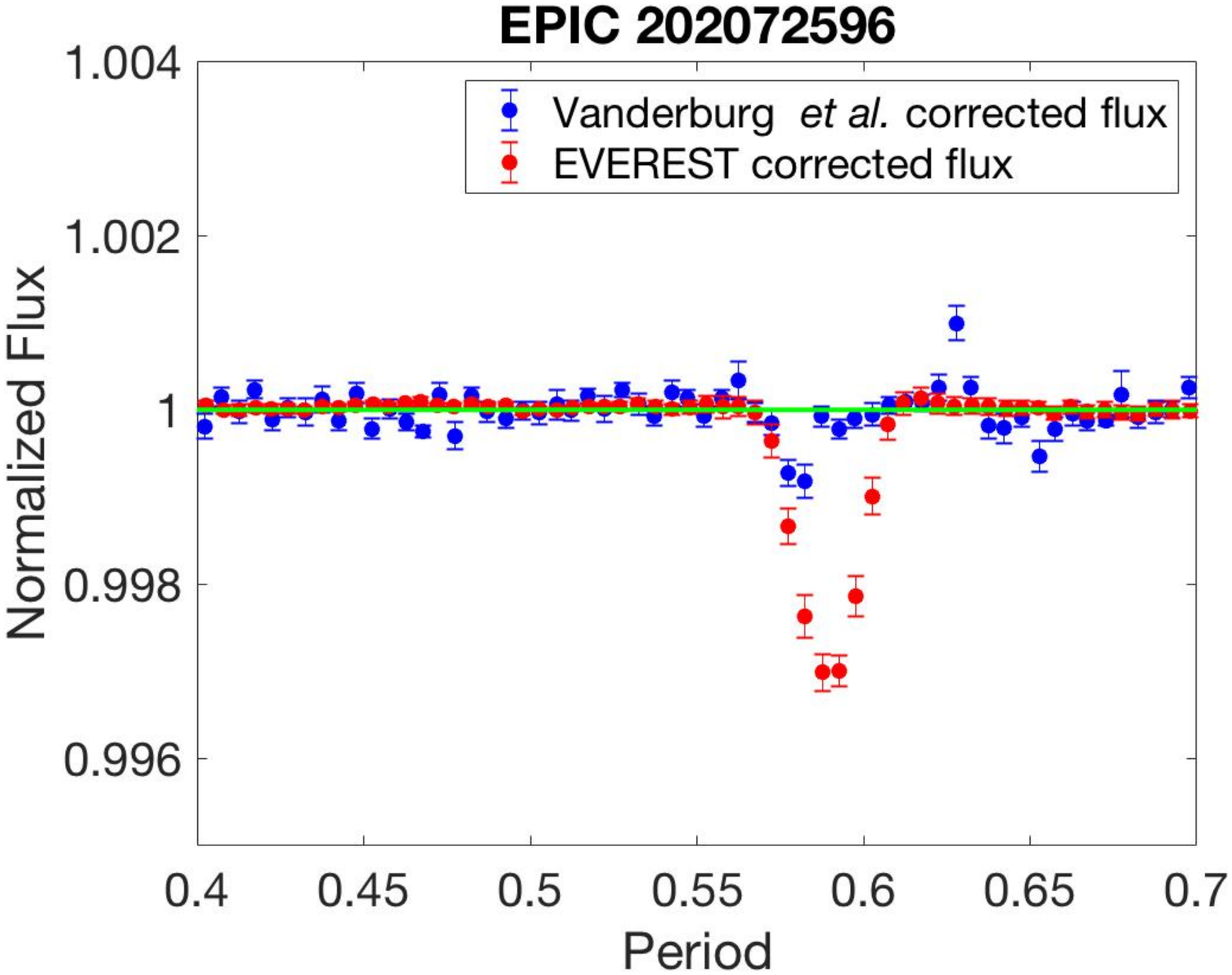}
  \caption{A comparison of the observed secondary eclipses between EVEREST and \cite{vanderburg} after the same post-pipeline detrending process. The green line at unity shows the average out-of-transit flux level. The flux is binned with bin size 0.005.}
  \label{Fig 4}
\end{figure}

We further confirmed the non-planetary nature of these objects by calculating their equilibrium and brightness temperatures. We calculated the theoretical equilibrium temperatures by taking a bond albedo of 0.3 in accordance with the low albedos generally favored for hot exoplanets \citep{cowan}. An exception is EPIC 203867512, which has a significantly longer orbital period than the other candidates (not in the temperature range covered by \citealp{cowan}).  Thus, we conservatively adopted an albedo of 0 for EPIC 203867512. We assumed uniform heat redistribution on all the planets, using the following formula from \cite{cowan}:

\begin{equation}
T_\textsc{p} = \textit{f}^\frac{1}{4}(1-A_{\textsc{B}})^\frac{1}{4}\frac{T_\textsc{s}}{\sqrt[]{a}_{*}}
\end{equation}

where $T_\textsc{s}$ is the stellar temperature reported in V16, $a_{*}$ is the transit parameter a/R$_{*}$ determined by equation 18 in \cite{seager} that is derived using only transit light curve observables and is independent of the host star mass, $\textit{f}$ is the factor related to the atmospheric circulation which we set to 1/4 for uniform temperature distribution, and $A_{\textsc{B}}$ is the bond albedo, which we set to 0.3 in each case except EPIC 203867512.

We then calculated the brightness temperature of the planet candidate from the observed depth of the secondary eclipse. We used the reported temperature of the host star from V16 in determining model atmosphere brightness temperatures over the range of wavelengths in the Kepler bandpass. We used the formula:

\begin{equation}
d = (\frac{R_\textsc{p}}{R_\textsc{s}})^2 \frac{\int{B_\textsc{$\lambda$}(T_\textsc{p})}K(\lambda)d\lambda}{\int{B_\textsc{$\lambda$}(T_\textsc{s})}K(\lambda)d\lambda}
\end{equation}

where $R_\textsc{p}$/$R_\textsc{s}$ is the ratio of planet to stellar radii, B$_\textsc{$\lambda$}(T_\textsc{s})$ is the intensity of the star using a model atmosphere \citep{kurucz} and the stellar temperature reported in V16, $K(\lambda)$ is the Kepler transmission function, and B$_\textsc{$\lambda$}(T_\textsc{p})$ is the Planck blackbody intensity of the planet where the temperature was varied until it reached a best fit value with the observed secondary depth, d, after integrating these functions over the wavelengths of the Kepler bandpass.

Equilibrium and brightness temperature values are reported in Table 1. Furthermore, we calculate the theoretical maximum observed brightness temperatures of these objects, by assuming that they are in fact BEBs, and thus taking the undiluted primary eclipse depth to be unity and recalculating the temperature from the appropriately scaled secondary eclipse depth using equation 2. These values are also reported in table 1.

It should also be noted that several of the systems we analyze have non-zero orbital eccentricities, because their secondary eclipses are displaced from phase 0.5 (see Figure~2). We calculate the ecos$\omega$ values for these systems using the eclipse offset from mid-phase value and equation 20 in \cite{pal}.

\begin{table*}
\begin{center}
\begin{supertabular}{c c c c c c c c c}
EPIC & $T_{Eq}$ & $T_{Obs}$ & $T_{max}$ & P & $R_p$/$R_{*}$ & d & ecos$\omega$\\ \hline
202071289&1002 $\pm$ 49&> 1591 $\substack{+20\\-24}$&4174&3.0459&0.1997&2750&0.0031\\ \hline
202071635&1032 $\pm$ 51&> 1431 $\substack{+47\\-65}$&3686&6.2719&0.3733&1500&0.1841\\ \hline
202072596&1670 $\pm$ 82&> 1767 $\substack{+70\\-97}$&5064&3.9792&0.2167&3100&0.1382\\ \hline
203867512&726 $\pm$ 36&> 1506 $\substack{+45\\-59}$&4765&28.4656&0.1642&1000&0.1219\\ \hline
205148699&1194 $\pm$ 59&> 1426 $\substack{+43\\-59}$&4202&4.3772&0.1744&1000&0.0823\\ \hline
206135267&1034 $\pm$ 51&> 1488 $\substack{+57\\-84}$&3437&2.5730&0.216&4000&0.0119\\ \hline
203753577&871 $\pm$ 43&> 1523 $\substack{+39\\-50}$&7983&3.4007&0.06863&260&0.0039\\ \hline
\end{supertabular}
\end{center}
\caption{System Parameters. $T_{Eq}$ is the equilibrium temperature of the planet in Kelvin, calculated from the transit parameters in equation 1, which assumes 0.3 Bond albedo (except for EPIC 203867512 which assumes 0 albedo) and uniform heat redistribution, $T_{Obs}$ is the observed temperature of the planet from secondary eclipse (in Kelvin) described in equation 2, $T_{max}$ is the theoretical maximum brightness temperature of the planet calculated by scaling the primary eclipse to unity, P is the period in days, $R_p$/$R_{*}$ is the planet to star radii ratio from the primary eclipse depth, d is the secondary eclipse depth in ppm, and ecos$\omega$ is the eccentricity with longitude of periastron argument derived from the secondary eclipse offset from midphase using equation 20 in \cite{pal}}.
\end{table*}

\section{Relevance to TESS}

The TESS mission \citep{ricker} has already begun to yield new planets transiting nearby bright stars \citep{huang, gandolfi, vanderspek}.  Vetting the large number of candidates identified by TESS will be a substantial task that can be facilitated by the best possible photometry.  We have investigated to what degree TESS candidates could be falsified based on the appearance of secondary eclipses.  We use the catalog of projected TESS planets calculated by \citet{barclay}, and for each system we replace the planet with a BEB.  That replacement assures that we are adding BEBs to the same types of stars where TESS will potentially discover planets.  We construct the BEBs by invoking the TRILEGAL galaxy simulation \citep{girardi} to produce a list of possible background stars that can be blended with each TESS planet-hosting star.  Since it would be impractical to run TRILEGAL for all of the individual planets in the Barclay simulation, we approximate the background stars using a grid of TRILEGAL runs.  We derive the background star properties (effective temperatures, radii, and TESS magnitudes) in 1-square degree fields at galactic longitudes 0, 30, 60, 90, 120, 150 and 180 degrees, and for each longitude, we run latitudes of 0, 30, and 60 degrees.  Also, we add a field at the north galactic pole.  Longitudes greater than 180-degrees, and negative galactic latitudes, are taken to have the same properties as their symmetric positions.  We used only a 0.001-square degree field for TRILEGAL at galactic center, since that provided more than sufficient stars in a reasonable computation time.  For each TESS planet-hosting star, we adopt the stellar background population from the spatially nearest point in our TRILEGAL grid.  

Given a TESS planet-hosting star, and a background stellar population, we construct a specific BEB as follows. We first choose a random star in the background population, under the condition that its brightness would be in an approximate range for consistency with the dilution factor required to match the observed transit of the planet candidate.  (When calculating the required dilution factor, we initially assume that the undiluted BEB will have a primary eclipse depth of unity.)  We then add a secondary component by choosing its mass randomly from a uniform distribution extending from the mass of the primary star down to 0.1\,$M_{\odot}$.  We adopt a main sequence evolutionary stage for the secondary star.  After choosing the stellar components of the BEB, we adjust the distance to that background system (slightly) in order to dilute the actual primary eclipse of the stellar system to equal the depth of the observed transit. We use Phoenix model atmospheres for the TRILEGAL stars when calculating eclipse depths.  This process also yields a depth for the secondary eclipse.  

We note that the mass function that describes the secondary components of eclipsing binaries has been uncertain in the literature (e.g., \citealp{halbwachs, soderhjelm, ducati, matson}), and can readily be affected by selection effects.  We use a constant mass function for the BEB secondaries as a reasonable compromise, given the uncertainty, but we also explored the limiting case of a minimum mass secondary (0.1 $M_{\odot}$) in all systems.  A minimum mass secondary is a worst-case when trying to distinguish BEBs from planets because the smallest and coolest secondary star produces secondary eclipses that are the hardest to detect, and are the closest to planets. 

\begin{figure}
  \includegraphics[width=\linewidth]{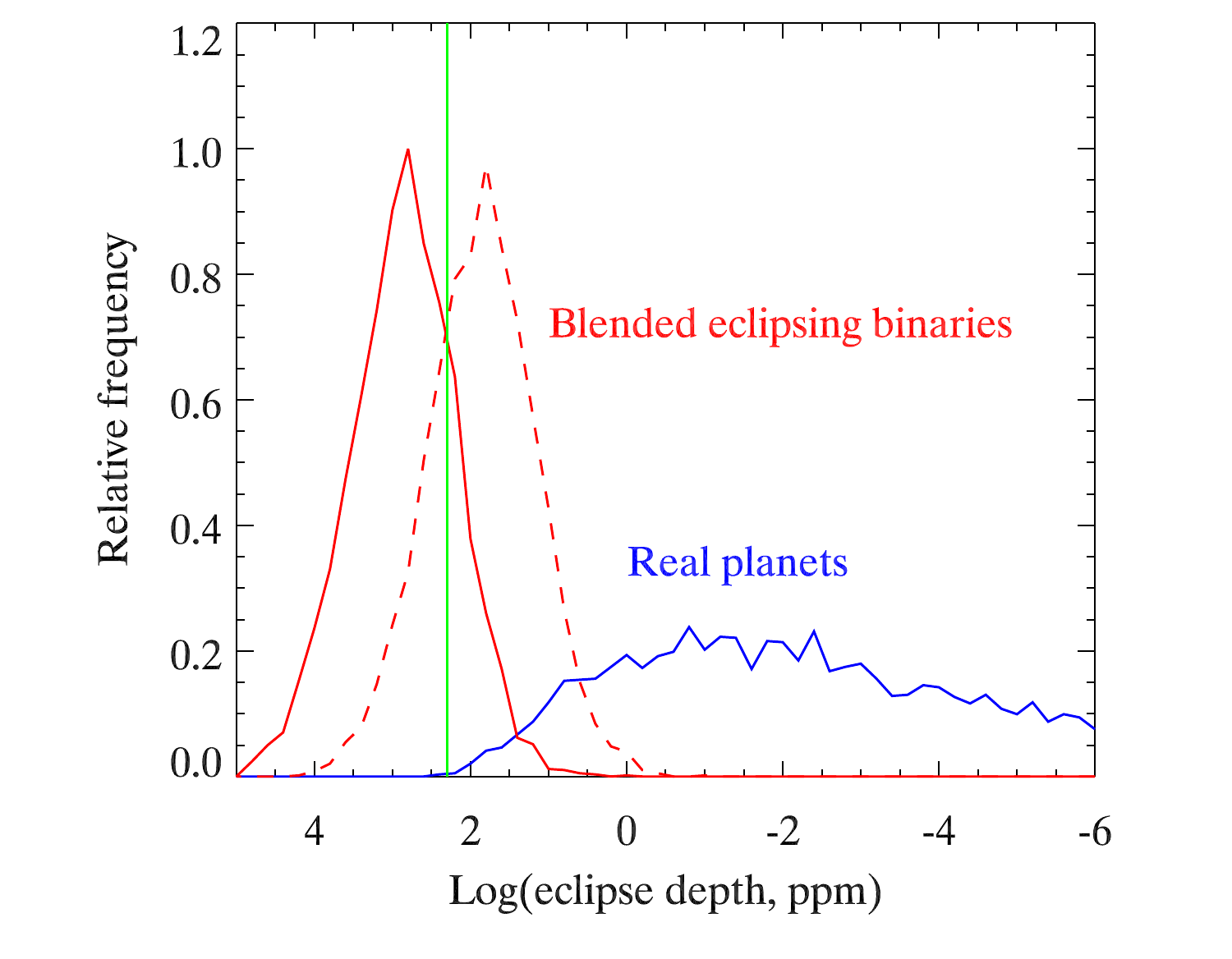}
  \caption{Distribution of secondary eclipse depths for blended eclipsing binaries (BEBs, in red) versus real planets in the TESS data (in blue).  We constructed BEBs using blended stars from TRILEGAL galaxy simulations (see text). The dashed red line shows the distribution expected for the case where every BEB has a minimum mass secondary star (0.1\,$M_{\odot}$).   Secondary eclipses from the BEBs will be detectable in many (but not all) cases using TESS photometry, whereas very few of the real planets will have detectable eclipses in TESS photometry, based on the catalog from \citet{barclay}.  The green line is the limiting eclipse depth for a $4\sigma$ detection when stacking all eclipses in a system having a TESS magnitude of 10, and median planet properties (e.g., eclipse duration) from the TESS alerts to date.}
  \label{Fig_5}
\end{figure}

Our results for TESS secondary eclipses are shown in Figure~\ref{Fig_5}.  That Figure shows the distribution of secondary eclipse depths for the catalog of simulated planets from \citet{barclay}.  We constructed that distribution by replacing each planet from \citet{barclay} with a BEB.  That replacement should not be interpreted as indicating anything about the rate of BEB occurrence.  Rather, we are comparing the relative depths of secondary eclipses produced by BEBs versus real planets for the same host stars.  As expected, the BEBs produce significantly stronger secondary eclipses, peaking near 200 ppm.  Real planets produce eclipse depths that are mostly negligible for TESS, the largest being $\sim 100$ ppm, but the peak of the broad distribution is near 0.01 ppm.  The Figure marks the $4\sigma$ detection limit for a system of TESS magnitude = 10, and virtually all eclipses of planets from \citet{barclay} are below that limit.  We also explicitly calculated the distribution of signal-to-noise for the eclipses of the planets in the \citet{barclay} catalog, and we find only 6 planets whose eclipses would be detectable above a $4\sigma$ threshold.  (That of course does not include the substantial number of currently known hot Jupiters whose eclipses may well be detectable in TESS photometry).  If the mass ratio in BEBs is represented by a uniform distribution, then Figure~\ref{Fig_5} indicates that about 2/3 their secondary eclipses are detectable in TESS photometry. In the worst-case when the BEB secondaries are all at minimum mass, then only about 1/3 of the BEB secondary eclipses are detectable by TESS.  But any detected secondary eclipse is unlikely to be a real planet, and confidence in that rejection can be enhanced using the brightness temperature analysis, as we have done here for K2 candidates.

Finally, we note that TESS should enable greater insight into the mass function in eclipsing binary stars, based on the substantial difference in the solid and dashed BEB distributions on Figure~\ref{Fig_5}.  However, inverting the secondary eclipse distribution to infer a mass distribution will require reliance on other statistical properties of binary stars, as well as correcting for the obvious selection effect: TESS will tend to discover such systems when the {\it primary} eclipse is deep, biasing the distribution toward systems with high mass primaries. 

\section{Results \& Discussion}
We determine that the following objects: EPIC 202071289, EPIC 202071635, EPIC 203867512, EPIC 205148699, EPIC 206135267, and EPIC 203753577 can be invalidated as planetary candidates and instead classified as eclipsing binaries, due to the clear presence of a large secondary eclipse, and the low equilibrium temperatures of the orbiting objects compared to the calculated brightness temperatures from the observed eclipses. For these systems, the temperature derived from the observed secondary eclipse depth is significantly greater than the theoretical equilibrium temperature given the temperature of the host star. This means that the objects are likely radiating their own light at a temperature inconsistent with being a planet. In the case of the one system, EPIC 202072596, where the equilibrium and observed temperatures overlap within 1$\sigma$, the significant depth of the secondary eclipse is  enough to warrant a strong suspicion that this system is an eclipsing binary, but does not provide a rigorous justification. The very high brightness temperatures of all of these planet candidates are more consistent with the brown dwarf range \citep{browndwarf}.

These detections are extracted exclusively from the improved photometry using the EVEREST pipeline. We show the secondary eclipses for all systems in Figure 2. We also calculate the theoretical maximum temperatures of these eclipsing binary objects, as well as EPIC 202072596, and estimate ecos$\omega$ values. The eccentricity of these systems, listed in Table 1, is yet another hint towards their non-planetary nature, as the eccentricity distribution of short period single planet candidates detected by Kepler in occultation has been shown to be tightly distributed around a zero mean \citep{eccentricity}, likely due to the short timescales of tidal circularization for these close-in planets candidates. However, one caveat must be mentioned. In the circumstance when a planet has a significant orbital eccentricity, and the radiative time scale is shorter than an orbital period, the planet could be hotter at secondary eclipse than the average temperature over the orbit.  An extreme case is a large eccentricity for a system that we view along the major axis of the orbit.  If the transit in that case occurs at apastron and the secondary eclipse at periastron, then the planet could indeed be significantly hotter than our calculated equilibrium temperature.  However, those are special and unlikely circumstances, hence we remain confident that the systems in Table~1 are indeed BEBs.

One system, EPIC 203753577, was originally reported in V16 to have a planetary radius of 5.57 Earth radii. The presence of a secondary eclipse in this system is noteworthy. Given the significant secondary eclipse, it is very likely a BEB, but suggests that other candidates with small radii might also have secondary eclipses belying their BEB nature that have been washed out in the out of transit flux variations.

Some or all of these systems have been previously examined in either \cite{vanderburg} or \cite{crossfield}, but neither explicitly demonstrated their improbability as planets solely from their photometric pipelines. We have done so here. We build upon their work because of our use of the EVEREST pipeline, which produced much better photometry that enabled us to clearly identify the deep secondary eclipses, and then calculate relevant parameters to disqualify these candidates. It is possible that there are also smaller undiscovered planet candidates whose primary transits were not initially flagged due to being washed out by the larger variance.  Some of those systems might now become visible using EVEREST photometry.

Some of the systems in this study were previously identified as probable detached eclipsing binaries from the K2 variability catalog (K2VARCAT), an archive maintained by the Space Telescope Science Institute that draws on the results of \cite{armstrong}. This speaks to the reliability of K2VARCAT as a tool for assigning value to potential candidates that require more careful vetting.

These six systems were relatively straightforward to demonstrate as eclipsing binaries, given the improved pipeline from EVEREST, a careful examination to mask the primary and secondary eclipses, and a smoothing of the out of transit flux. There are potentially many more systems in the list of planetary candidates from K2 that are also eclipsing binaries whose secondary eclipses were not immediately visible in the raw or preliminarily corrected data. A further examination of systems with a smaller planet to star size ratio might reveal these similar false positive conditions, or even systems of confirmed candidates that have secondary eclipses yet to be discerned from the data. The systems examined here comprise only candidate results from campaigns 0 - 5. When EVEREST data becomes available for the last few campaigns, and for Kepler and TESS, this same analysis will be possible for a great many more systems.

For observations with TESS, we demonstrate that most secondary eclipses from real planets will not be detectable.  Assuming that the secondary-to-primary mass ratio in BEBs is uniformly distributed, we project that 2/3 of BEB secondary eclipses will be detected by TESS, but those BEBs can be invalidated as planet candidates using the brightness temperature analysis demonstrated here.

\bibliography{references.bib}

\end{document}